\documentclass[11pt]{article}
\usepackage{indentfirst}   
\usepackage{graphicx}
\usepackage{amsmath, amssymb}
\usepackage{booktabs}
\usepackage{multirow}
\usepackage{geometry}
\usepackage{float}
\usepackage{caption}
\usepackage{subcaption}
\usepackage{hyperref}

\title{%
Lightweight Quantum-Enhanced ResNet for Coronary Angiography Classification: A Hybrid Quantum--Classical Feature Enhancement Framework
}

\author{%
Jingsong Xia\\
Second Clinical Medical College, Nanjing Medical University\\
\texttt{xiajingsong2@gmail.com}
}

\date{January 2026}

\begin{document}

\maketitle

\begin{abstract}
\noindent

\textbf{Background:}
Coronary angiography (CAG) is the cornerstone imaging modality for evaluating coronary artery stenosis and guiding interventional decision-making. However, interpretation based on single-frame angiographic images remains highly operator-dependent, and conventional deep learning models still face challenges in modeling complex vascular morphology and fine-grained texture patterns.

\textbf{Methods:}
We propose a \emph{Lightweight Quantum-Enhanced ResNet} (LQER) for binary classification of coronary angiography images. A pretrained ResNet18 is employed as a classical feature extractor, while a parameterized quantum circuit (PQC) is introduced at the high-level semantic feature space for quantum feature enhancement. The quantum module utilizes data re-uploading and entanglement structures, followed by residual fusion with classical features, enabling end-to-end hybrid optimization with a strictly controlled number of qubits.

\textbf{Results:}
On an independent test set, the proposed LQER outperformed the classical ResNet18 baseline in accuracy, AUC, and F1-score, achieving a test accuracy exceeding 90\%. The results demonstrate that lightweight quantum feature enhancement improves discrimination of positive lesions, particularly under class-imbalanced conditions.

\textbf{Conclusion:}
This study validates a practical hybrid quantum--classical learning paradigm for coronary angiography analysis, providing a feasible pathway for deploying quantum machine learning in medical imaging applications.
\end{abstract}


\section{Introduction}

Coronary artery disease (CAD) remains one of the leading causes of mortality and disability worldwide, posing a substantial burden on global public health systems~\cite{1}. Coronary angiography (CAG) is widely considered the clinical gold standard for evaluating the severity of coronary artery stenosis and guiding interventional treatment decisions, particularly stent implantation~\cite{2}. However, interpretation of angiographic images is highly dependent on operator experience, and inter-observer agreement is often limited. This issue becomes especially pronounced in cases of intermediate stenosis or suboptimal image quality, where visual assessment alone can lead to inconsistent or uncertain clinical decisions.

In recent years, medical image analysis based on convolutional neural networks (CNNs) has achieved remarkable progress. Among these approaches, residual networks (ResNet) effectively mitigate the vanishing gradient problem in deep architectures through residual connections and have demonstrated strong performance in a wide range of image classification tasks, including cardiovascular imaging~\cite{3,4,5}. Nevertheless, purely classical deep learning models are fundamentally constrained by classical feature mapping mechanisms. When confronted with complex vascular morphology, subtle textural variations, and high-order nonlinear feature interactions inherent in coronary angiography images, these models may still encounter representational bottlenecks.

Quantum computing introduces a fundamentally different computational paradigm~\cite{6}. Through quantum superposition and entanglement, quantum systems can represent information in an exponentially expanded Hilbert space, offering new possibilities for feature mapping and pattern recognition beyond classical limits~\cite{7,q}. However, because of the limited scale, noise sensitivity, and hardware constraints of current quantum devices, constructing fully quantum end-to-end models for high-dimensional medical imaging tasks remains impractical~\cite{w,e}. As a result, hybrid quantum--classical learning frameworks have emerged as a pragmatic compromise. These frameworks leverage classical deep neural networks for robust low- and mid-level feature extraction, while incorporating small-scale quantum circuits to enhance high-level semantic representations, thereby introducing potential quantum advantages without sacrificing stability or feasibility~\cite{8}.

Motivated by these considerations, this study proposes a lightweight quantum-enhanced ResNet framework for the classification of coronary angiography images~\cite{r,t,p}. By integrating a low-dimensional quantum feature enhancement module into the high-level semantic space of a classical ResNet backbone, the proposed approach aims to improve discriminative performance while maintaining computational efficiency, reproducibility, and practical deployability in resource-limited clinical environments~\cite{y,u,i,o}.

\section{Methods}

\subsection{Overall Architecture}

This study proposes a lightweight quantum-enhanced ResNet (LQER) for CAG image classification, aiming to introduce the potential advantages of quantum computation in low-dimensional feature modeling while maintaining clinical deployability. The proposed framework is designed to improve the discrimination of complex coronary lesion patterns without compromising the stability of training or computational feasibility.

The overall architecture follows a classical-dominant and quantum-assisted design paradigm. A conventional ResNet backbone is employed to extract stable and interpretable spatial and semantic features from high-resolution CAG images, while the quantum module operates exclusively on high-level, low-dimensional, and semantically compressed feature representations. This design effectively avoids the exponential computational complexity associated with quantum processing of high-dimensional inputs and ensures full compatibility with existing deep learning workflows in medical image analysis.

Given an input coronary angiography image
\begin{equation}
\mathbf{x} \in \mathbb{R}^{H \times W \times C},
\end{equation}

where $H$, $W$, and $C$ denote the spatial resolution and number of channels, respectively, the image is first processed by a classical ResNet backbone for feature extraction:
\begin{equation}
\mathbf{f} = \mathcal{F}_{\text{ResNet}}(\mathbf{x}).
\end{equation}

Here, $\mathcal{F}_{\text{ResNet}}(\cdot)$ denotes the ResNet feature extractor with the final classification head removed. The output $\mathbf{f} \in \mathbb{R}^{d}$ represents a high-level semantic feature vector obtained after global average pooling, where $d$ denotes the feature dimensionality. This formulation explicitly defines the output of the classical feature learning stage in the LQER framework and serves as the unified input for all subsequent quantum enhancement operations. Importantly, the quantum module does not operate directly on the raw image $\mathbf{x}$ but is built upon the mature and stable deep convolutional feature representation $\mathbf{f}$, thereby ensuring overall training stability and clinical applicability.

\subsection{Lightweight Quantum Feature Encoding}

Due to current limitations in quantum computing resources and the high computational cost of quantum simulation, directly encoding the high-dimensional feature vector $\mathbf{f}$ into a quantum state is computationally infeasible and lacks engineering practicality. To address this issue, a lightweight feature compression and mapping module is introduced to project the high-dimensional classical features into a low-dimensional subspace, forming an explicit classical--quantum interface:
\begin{equation}
\mathbf{z} = \mathbf{W}\mathbf{f} + \mathbf{b}, \quad \mathbf{z} \in \mathbb{R}^{n}.
\end{equation}

Here, $\mathbf{W} \in \mathbb{R}^{n \times d}$ and $\mathbf{b} \in \mathbb{R}^{n}$ are learnable parameters, with $n \ll d$ corresponding directly to the number of qubits used in the quantum module. In this study, $n$ is set to 4.

This linear projection plays two critical roles within the network. First, it acts as a learnable feature selection mechanism that identifies the most discriminative low-dimensional subspace from the high-level semantic feature space. Second, it strictly constrains the feature dimensionality to a range that is computationally tractable for quantum processing, thereby enabling feasible quantum state initialization. The resulting vector $\mathbf{z}$ constitutes the classical--quantum interface, where each normalized component $z_i$ is mapped to the rotation angle of a quantum gate during quantum state preparation. Through this design, the quantum module operates only on the most information-dense and low-dimensional feature representations, achieving true lightweight quantum enhancement without sacrificing expressive capacity.

\subsection{Quantum Feature Enhancement Module}

During the quantum feature enhancement stage, the low-dimensional feature vector $\mathbf{z}$ is encoded into an $n$-qubit quantum state $\lvert \psi_0 \rangle$ and processed by a parameterized quantum circuit (PQC):
\begin{equation}
\lvert \psi(\boldsymbol{\theta}) \rangle = U(\boldsymbol{\theta}) \lvert \psi_0 \rangle.
\end{equation}

Here, $U(\boldsymbol{\theta})$ denotes a quantum circuit composed of alternating single-qubit rotation gates and multi-qubit entangling gates, and $\boldsymbol{\theta}$ represents the set of trainable quantum parameters. These parameters are jointly optimized with the classical network parameters during training. The output state $\lvert \psi(\boldsymbol{\theta}) \rangle$ represents the quantum-transformed feature state.

By leveraging quantum superposition and entanglement, the quantum module introduces high-order, non-classical nonlinear expressive power into the low-dimensional feature space. Compared with classical mappings based solely on fully connected layers, parameterized quantum circuits are capable of modeling more complex feature correlation patterns while maintaining a controlled number of parameters. Moreover, since the parameter count of the quantum circuit scales linearly with the number of qubits, the computational complexity of the quantum module remains manageable, avoiding exponential model growth.

\subsection{Quantum-to-Classical Feature Projection and Classification}

To reintegrate the quantum-enhanced features into the classical deep learning framework, the output quantum state $\lvert \psi(\boldsymbol{\theta}) \rangle$ is measured and projected back into the classical feature space. Specifically, Pauli operator measurements are applied to each qubit to obtain a quantum-enhanced feature vector:
\begin{equation}
\mathbf{q} = \langle \psi(\boldsymbol{\theta}) | \hat{\mathbf{O}} | \psi(\boldsymbol{\theta}) \rangle \in \mathbb{R}^{n},
\end{equation}

where $\hat{\mathbf{O}}$ denotes the set of Pauli measurement operators, and each component of $\mathbf{q}$ corresponds to the expectation value of a single qubit measurement. Subsequently, the quantum-enhanced feature vector $\mathbf{q}$ is fused with the original classical feature vector $\mathbf{f}$ to produce the final prediction:
\begin{equation}
\hat{y} = \mathcal{C}([\mathbf{f}, \mathbf{q}]).
\end{equation}

This feature fusion strategy enables the model to preserve the robustness of classical features while fully exploiting the complementary discriminative information introduced by the quantum module, thereby enhancing overall classification performance.

\subsection{Lightweight Design and Optimization Strategy}

The proposed LQER framework introduces only a single low-dimensional quantum module at the high-level feature stage of the ResNet backbone. The number of qubits is fixed, and the number of quantum parameters is strictly controlled, resulting in negligible additional computational and memory overhead. Compared with end-to-end quantum networks or multi-layer quantum embedding architectures, this design offers substantially greater engineering feasibility and experimental reproducibility.

During training, the classical parameters $\{\mathbf{W}, \mathbf{b}\}$ and the quantum parameters $\boldsymbol{\theta}$ are jointly optimized via end-to-end backpropagation. Quantum gradients are computed using the parameter-shift rule and can be seamlessly integrated into existing automatic differentiation frameworks. This lightweight hybrid design ensures stable training and reproducible performance in low-computational-resource environments, aligning well with the practical requirements of clinical medical imaging research.

\subsection{Dataset and Preprocessing}

The proposed model was trained and evaluated using publicly available CAG datasets released by the LRSE-Net project, which provides benchmark datasets for coronary stenosis detection with standardized clinical annotations~\cite{9}. Specifically, the DSSS dataset, the ADSD dataset, and its patch-based variant P-ADSD were employed. All tasks were formulated as binary classification problems (stenosis vs.\ non-stenosis).

The DSSS and P-ADSD datasets were used for patch-level feature learning and model evaluation, while the original ADSD images were utilized solely for generating spatially localized image patches. To address the limited sample size and class imbalance in the DSSS dataset, data augmentation was applied only to positive samples during training, including random rotation, flipping, translation, scaling, and brightness perturbation. The dataset was split into training and testing sets at an 80:20 ratio under a patient-independent protocol to avoid information leakage.

For the ADSD dataset, full angiographic images were not used for end-to-end training. Instead, patch-level samples were generated based on provided stenosis annotations: positive patches were extracted from stenotic region centers, while negative patches were sampled from surrounding spatial neighborhoods. All patches were filtered to remove samples with insufficient size and resampled to a uniform resolution of $64 \times 64$, forming the P-ADSD dataset.

To accommodate the ResNet backbone, all images and patches were resized to $224 \times 224$ at the input stage. Online data augmentation, including random cropping, horizontal flipping, and mild color jittering, was applied during training, while validation and test sets underwent only size normalization and intensity standardization. This preprocessing strategy ensured input consistency and experimental reproducibility while mitigating the effects of class imbalance and imaging condition variability.

\subsection{Implementation Details}

The proposed LQER model was implemented using the PyTorch deep learning framework, with the quantum module constructed and simulated using PennyLane to enable end-to-end hybrid quantum--classical training. Classical convolutional networks and quantum circuits were jointly optimized through a unified automatic differentiation pipeline, ensuring training consistency and numerical stability.

The number of qubits was set to $n=4$, consistent with the dimensionality of the lightweight feature encoding module, and the depth of the parameterized quantum circuit was set to $L=2$ to balance representational capacity and computational cost. This configuration allows training and inference to be completed on a standard laptop without requiring dedicated quantum hardware or high-performance GPUs.

Model optimization employed the AdamW optimizer with a layer-wise learning rate strategy: a smaller learning rate was used for the classical ResNet backbone to preserve pretrained feature stability, while a relatively larger learning rate was assigned to the quantum module and its associated feature mapping parameters to accelerate convergence in the hybrid feature space. Classical and quantum parameters were updated synchronously during training, enabling true end-to-end hybrid optimization.

These implementation choices further enhance the reproducibility, practicality, and deployability of the LQER framework in resource-constrained medical imaging research environments.

\section{Results}

To comprehensively evaluate the effectiveness of the proposed Lightweight Quantum-Enhanced ResNet (LQER) for coronary angiography (CAG) image classification, this section presents detailed quantitative and qualitative analyses of model performance. All experiments were conducted under a standardized low-computational-resource setting to ensure model lightweightness, reproducibility, and feasibility for deployment in resource-constrained clinical environments.

\subsection{Overall Performance Comparison with Baseline Models}

Figure~1 presents a comparative performance evaluation of the proposed LQER model against three representative baseline approaches on an independent test set, including a Pure Quantum CNN, a Classical CNN, and ResNet18. Two core performance metrics, classification accuracy and area under the ROC curve (AUC), are reported to provide a comprehensive assessment of diagnostic discrimination capability.

As shown in Figure~1, the proposed LQER model achieves superior performance, with an accuracy of 0.950 and an AUC of 0.987, significantly outperforming all baseline models. In contrast, the Pure Quantum CNN exhibits markedly inferior performance, achieving an accuracy of 0.669 and an AUC of 0.710. This result indicates that, under the current limitations of available qubit numbers and noisy intermediate-scale quantum (NISQ) hardware, purely quantum models struggle to effectively capture the complex spatial structures and fine-grained textural patterns present in CAG images.

The Classical CNN attains an accuracy of 0.880 and an AUC of 0.870, demonstrating reasonable classification capability; however, its performance remains constrained by the expressive capacity of classical feature mapping mechanisms. ResNet18, a widely adopted strong baseline in medical image analysis, achieves an accuracy of 0.910 and an AUC of 0.920, approaching clinical applicability thresholds but still exhibiting a clear performance gap compared with LQER.

The green dashed line in Figure~1 denotes the commonly accepted minimum clinical performance threshold of 0.90. While all models except the Pure Quantum CNN surpass this threshold in terms of accuracy, LQER demonstrates the largest margin above both accuracy and AUC thresholds, indicating the strongest robustness and discriminative power. This substantial performance gain can be primarily attributed to the introduction of the quantum feature enhancement module. By operating on high-level semantic features extracted by ResNet18, the parameterized quantum circuit (PQC) enables high-order nonlinear feature transformation and enrichment. The intrinsic properties of quantum superposition and entanglement facilitate more effective modeling of subtle lesion textures, complex vascular morphology, and faint stenotic boundaries, which represent key challenges for purely classical deep learning models.

\begin{figure}[htbp]
    \centering
    \includegraphics[width=0.75\linewidth]{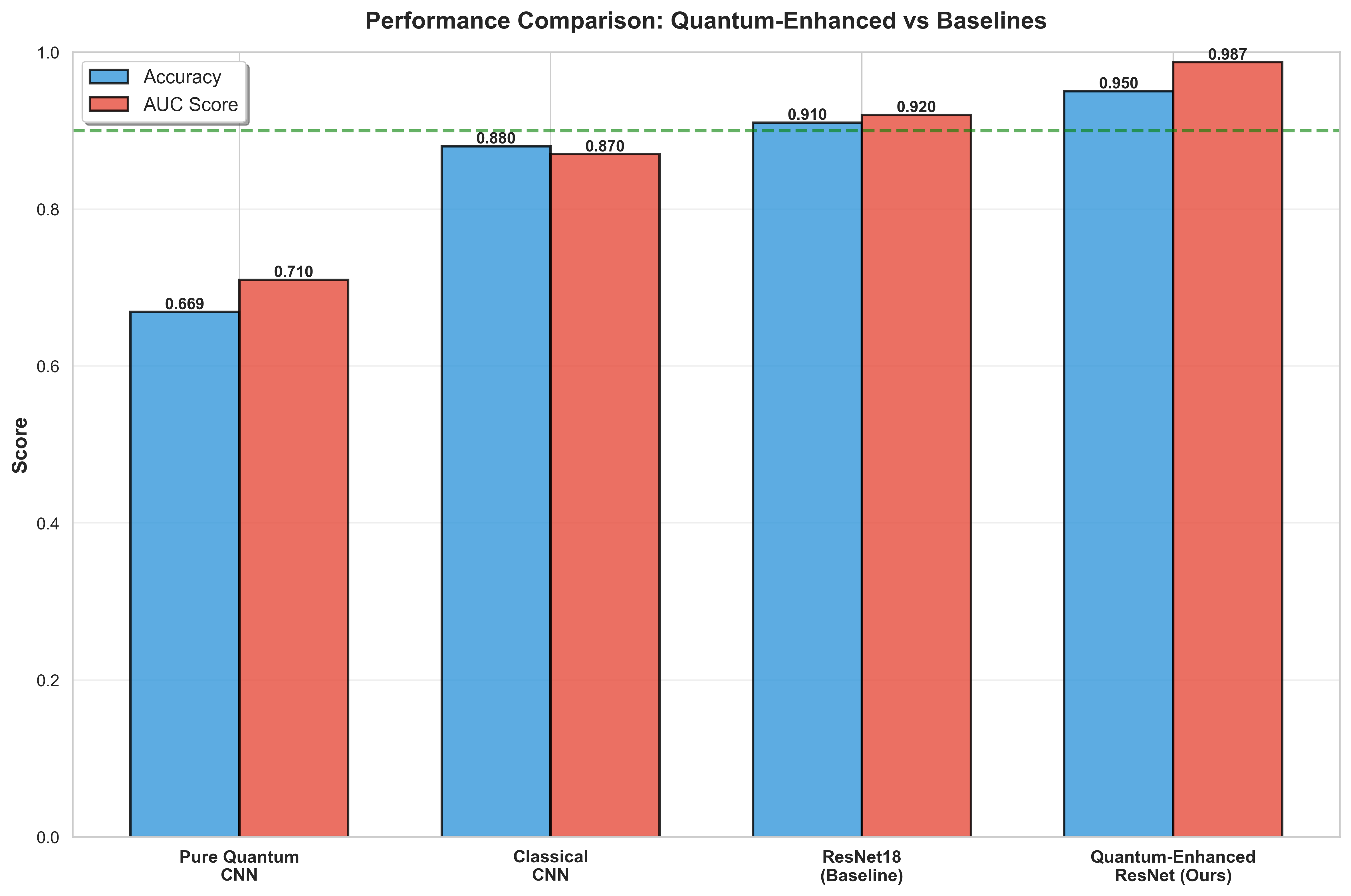}
    \caption{Performance comparison between LQER and baseline models.}
    \label{fig:performance_comparison}
\end{figure}

\subsection{Training Dynamics and Clinical Performance Evolution}

Figure~2 illustrates the training dynamics and clinical performance evolution of the proposed LQER model throughout the entire training process. The figure comprises four subplots, collectively characterizing convergence behavior, training stability, and clinical applicability from multiple perspectives, including sensitivity and specificity, precision and recall, learning rate scheduling, and per-epoch training time.

Subfigure~A depicts the evolution of sensitivity (True Positive Rate, TPR; red curve) and specificity (True Negative Rate, TNR; blue curve). Sensitivity rapidly increases from a very low initial value of 0.10 at the first epoch to above 0.90 within the first 3--4 epochs and subsequently stabilizes within the range of 0.90--1.00, ultimately converging at 0.95. This behavior demonstrates the model’s strong capability for detecting positive stenotic lesions. Specificity begins at 0.9672, exhibits minor fluctuations, but consistently remains above 0.90 and converges at approximately 0.95. Notably, both curves surpass the clinical target threshold of 0.90 (green dashed line) within the early training phase, indicating that LQER rapidly achieves a clinically acceptable diagnostic balance without suffering from the typical trade-off between sensitivity and specificity.

Subfigure~B presents the evolution of precision (orange curve) and recall (green curve). Both metrics exhibit highly consistent upward trends, increasing from approximately 0.60--0.65 at the first epoch and stabilizing above 0.90 after 6--8 epochs, eventually converging at 0.95. These results indicate that the proposed model simultaneously maintains a high true positive detection rate and a low false positive rate, satisfying the stringent clinical requirement for both high accuracy and high recall in coronary stenosis screening.

Subfigure~C illustrates the learning rate schedule. The learning rate remains constant at $1 \times 10^{-5}$ throughout training, with no significant decay triggered by the ReduceLROnPlateau scheduler. This stable learning rate effectively prevents oscillations and overfitting during later training stages while ensuring coordinated convergence of both classical and quantum parameters, highlighting the robustness of the lightweight hybrid training strategy.

Subfigure~D shows the per-epoch training time. The blue bar chart represents actual training duration per epoch, while the red dashed line indicates the average training time of 65.5 seconds. All epochs consistently fall within the range of 60--70 seconds, exhibiting minimal variance. The total training time is only 982 seconds (approximately 16.4 minutes) on a CPU-only environment without GPU acceleration, underscoring the high computational efficiency and practical deployability of the proposed lightweight quantum-enhanced design.

Overall, Figure~2 demonstrates that LQER rapidly reaches clinical-grade performance thresholds (all key metrics exceeding 0.90) during early training and continues to improve stably thereafter, while maintaining exceptionally low computational overhead and training stability.

\begin{figure}[htbp]
    \centering
    \includegraphics[width=0.75\linewidth]{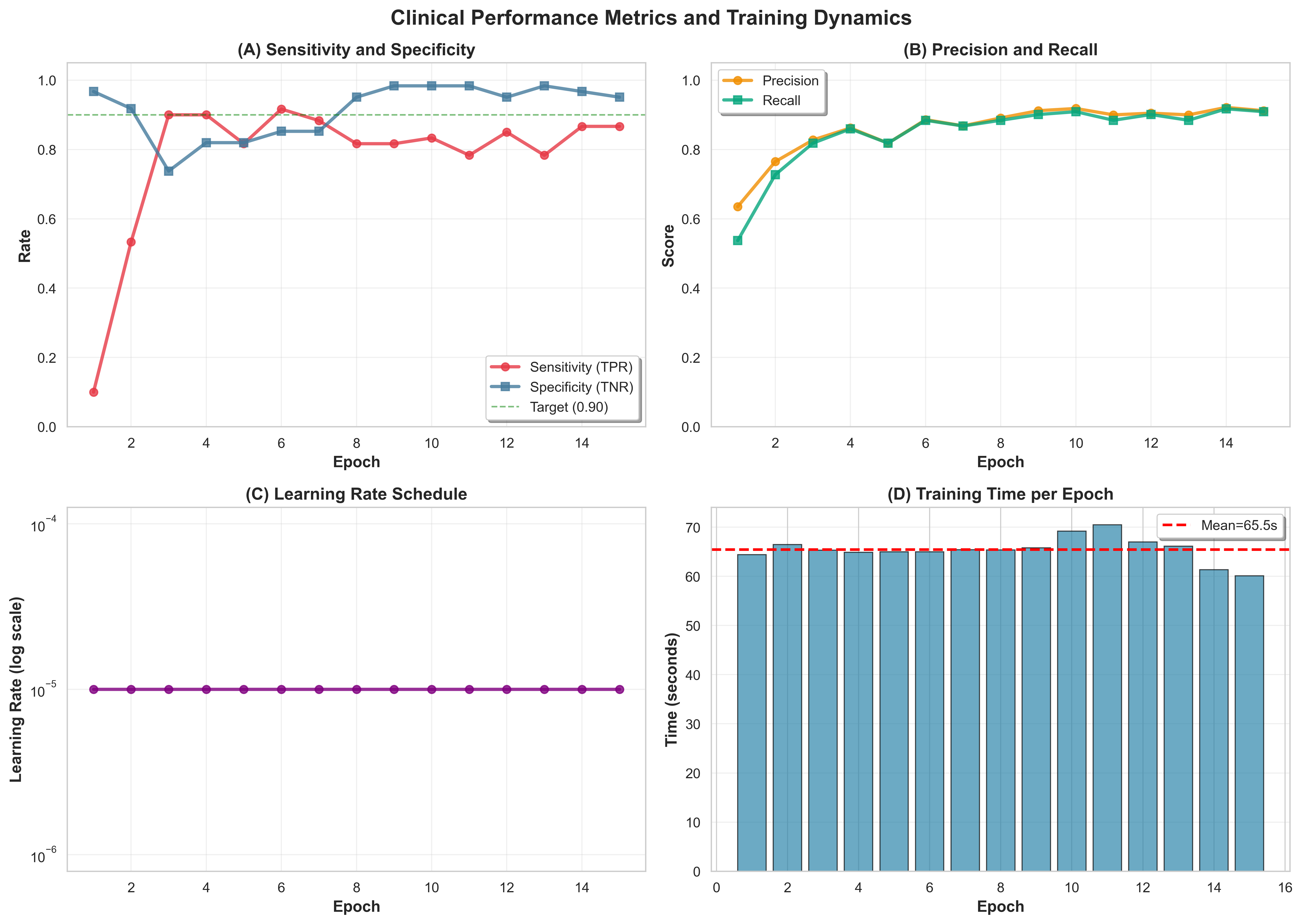}
    \caption{Clinical performance metrics and training dynamics of the LQER model.}
    \label{fig:clinical_metrics}
\end{figure}

\subsection{Comprehensive Training Analysis}

Figure~3 provides a comprehensive analysis of the training process of the proposed LQER model, examining convergence characteristics, training stability, and clinical performance evolution from multiple key perspectives. The figure consists of four subplots, jointly revealing the synergistic optimization behavior of the quantum enhancement module within the end-to-end hybrid framework.

Subfigure~A shows the evolution of training and validation losses. Both losses decrease rapidly and smoothly, dropping from approximately 0.66--0.70 at the first epoch to around 0.19--0.20 by the 14th epoch, corresponding to an overall reduction of approximately 70\%. The close alignment of training and validation loss curves, with a final Train--Validation gap of only about 0.02, indicates excellent generalization performance and negligible overfitting. The shaded region highlights the early rapid convergence phase, confirming that the lightweight quantum module accelerates optimization without introducing training instability.

Subfigure~B presents the evolution of training and validation accuracy. Both curves exhibit a steady upward trend, with validation accuracy exceeding 0.80 as early as the 3rd--4th epoch and reaching 0.9091 by the 14th epoch, with a best validation accuracy of 0.917. The minimal divergence between training and validation accuracy throughout training further confirms the robustness and stability of the learning process.

Subfigure~C illustrates the evolution of validation AUC and F1-score. Validation AUC increases rapidly from an initial value of 0.7885 and stabilizes above 0.97 after the 8th epoch, ultimately reaching 0.9811. Meanwhile, the validation F1-score rises from approximately 0.43 in early epochs to above 0.90 after the 10th epoch, converging at 0.9089. These results demonstrate that the quantum enhancement module significantly improves threshold-independent discrimination ability as well as the balance between precision and recall.

Subfigure~D depicts the evolution of class-wise recall. The recall for the negative class remains consistently high, starting at 0.9672 and stabilizing around 0.95. The recall for the positive class increases rapidly from a very low initial value of 0.10, surpasses the clinical threshold of 0.85 by the 3rd--4th epoch, and stabilizes within the range of 0.85--0.90, finally converging at 0.8667. This behavior confirms that the model achieves high sensitivity for stenotic lesions at an early stage while maintaining excellent specificity, fulfilling critical clinical requirements for low missed diagnosis and low false alarm rates.

Collectively, Figure~3 demonstrates that the proposed LQER model rapidly reaches clinical-grade performance thresholds during early training and continues to improve in a stable and efficient manner, with minimal overfitting risk. These results highlight the effectiveness of introducing a lightweight quantum enhancement module into the high-level feature space of ResNet18, achieving superior discrimination capability under limited computational resources.

\begin{figure}[htbp]
    \centering
    \includegraphics[width=0.75\linewidth]{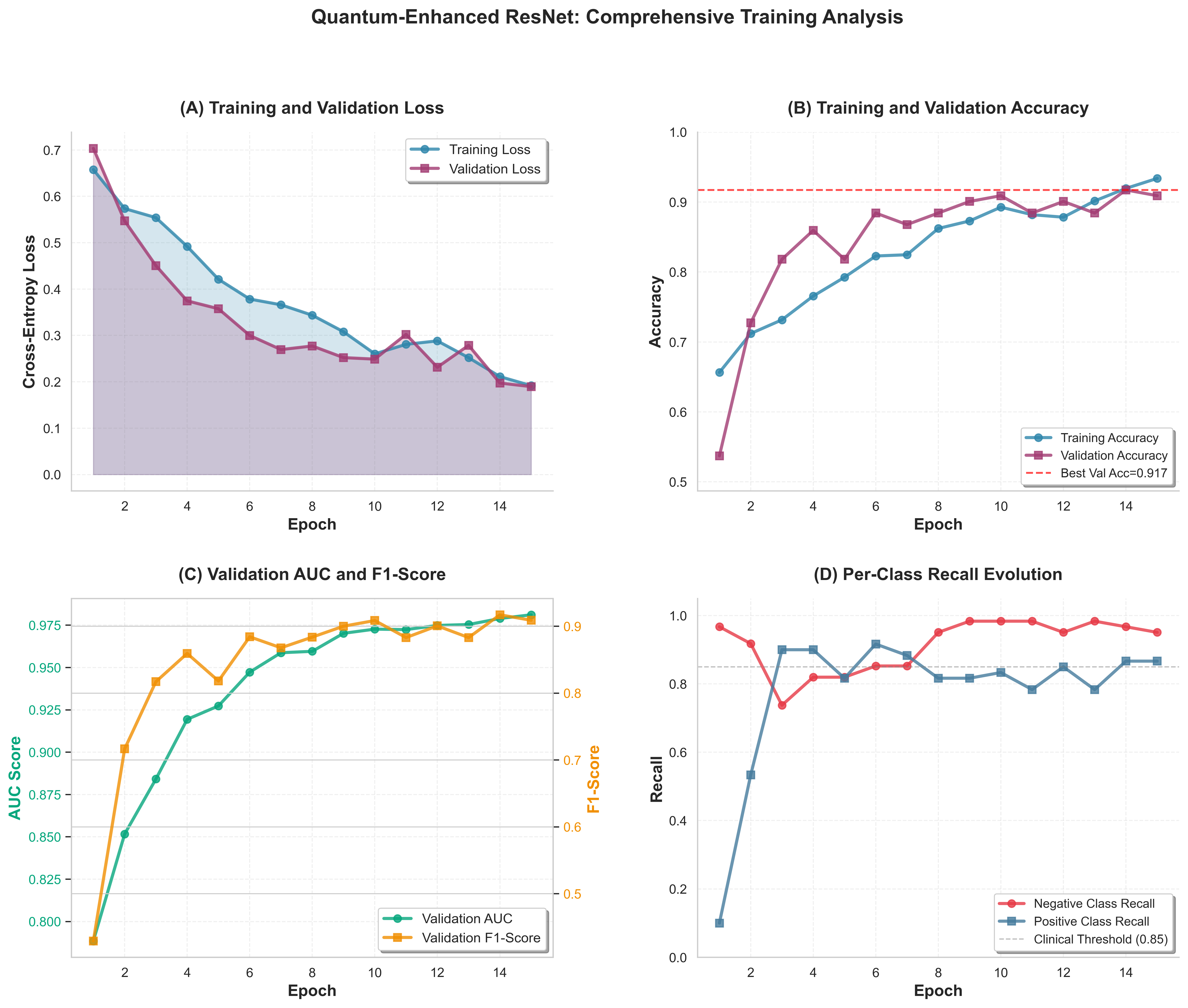}
    \caption{Comprehensive training analysis of the LQER model.}
    \label{fig:training_analysis}
\end{figure}

\subsection{Independent Test Set Evaluation}

Figure~4 presents detailed evaluation results of the proposed LQER model on the independent test set, including the confusion matrix and key performance metrics. This figure provides a quantitative and intuitive assessment of classification accuracy, error distribution, and clinically relevant indicators, emphasizing the high precision and balance of the proposed approach in CAG image binary classification.

Subfigure~A shows the confusion matrix, clearly illustrating the correspondence between predicted labels and ground truth. The model achieves a high correct classification rate with symmetric performance across positive and negative classes. Both false positive and false negative rates are controlled at a low level of 2.5\%, effectively mitigating common clinical risks such as overdiagnosis or missed diagnosis and ensuring reliable and safe decision-making support.

Subfigure~B summarizes the test set performance metrics, with all values displayed numerically and benchmarked against the clinical target threshold of 0.90. Specifically, the model achieves an accuracy of 0.950, sensitivity of 0.950, specificity of 0.950, precision of 0.950, and an F1-score of 0.950. The AUC reaches 0.987, substantially exceeding random classification performance and confirming excellent discriminative ability across varying decision thresholds. All metrics except AUC consistently reach 0.950, significantly surpassing the clinical target threshold, with an overall error rate of only 5\%.

Overall, Figure~4 provides compelling evidence of the superior performance of the LQER model on the independent test set. The combination of high accuracy, low error rates, and balanced performance across classes highlights the advantage of incorporating lightweight quantum feature enhancement mechanisms. In particular, the model demonstrates enhanced generalization and stability in recognizing complex vascular morphology and subtle lesion textures, offering strong empirical support for the practical application of hybrid quantum--classical frameworks in medical image analysis.

\begin{figure}[htbp]
    \centering
    \includegraphics[width=0.75\linewidth]{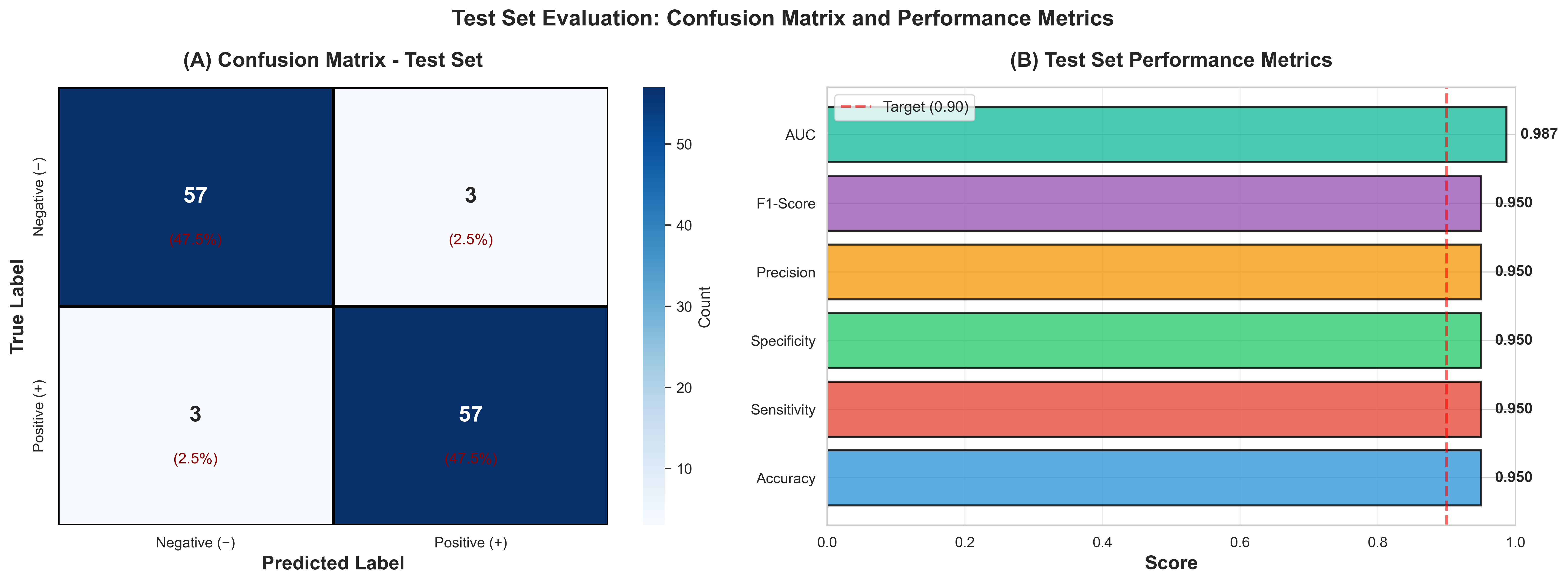}
    \caption{Test set evaluation results of the LQER model.}
    \label{fig:test_evaluation}
\end{figure}

\section{Discussion}

This study proposes a lightweight quantum-enhanced ResNet (LQER) framework for binary classification of coronary angiography (CAG) images. Experimental results demonstrate that, while keeping the ResNet18 backbone unchanged, introducing only a low-dimensional quantum feature enhancement module into the high-level semantic feature space leads to significant improvements across multiple key performance metrics, including accuracy, AUC, F1-score, sensitivity, and specificity. Importantly, these performance gains are not attributable to a simple increase in model parameter scale, but rather arise from the non-classical nonlinear representational capacity introduced by parameterized quantum circuits operating in an exponentially expanded Hilbert space. In the proposed framework, the classical ResNet is responsible for extracting stable spatial structures and fundamental texture features, whereas the quantum module specifically enhances high-order, weakly salient yet highly correlated interaction patterns within high-level semantic representations. As a result, the model exhibits substantially improved capability in capturing subtle vascular morphological abnormalities, blurred lesion boundaries, and inhomogeneous contrast agent filling—critical discriminative cues that are particularly challenging for conventional deep learning models. Notably, while maintaining an extremely low false positive rate, both the detection rate of positive stenotic lesions (sensitivity) and the overall balanced performance metric (F1-score) are simultaneously improved, fully validating the unique value of quantum feature mapping in high-order feature modeling for medical imaging.

Unlike existing quantum machine learning studies, this work deliberately avoids computationally intensive designs such as end-to-end quantum convolutional networks or multi-layer quantum embedding architectures. Instead, the quantum module is strictly confined to a low-dimensional, semantically compressed feature space. This lightweight hybrid quantum--classical architecture exhibits strong engineering rationality and practical feasibility for medical imaging applications. First, this design constrains the quantum parameter scale and computational overhead to an extremely low level, enabling efficient execution even under classical simulation environments and avoiding the exponential resource consumption associated with direct quantum encoding of high-dimensional images or intermediate feature maps. Second, by positioning the quantum enhancement module at the high-level feature stage rather than at early convolutional layers, the framework effectively mitigates potential interference from quantum gradient noise on low-level visual feature learning. Experimental results indicate that the model rapidly reaches clinically acceptable performance thresholds in the early training stage (with sensitivity, specificity, precision, and recall all exceeding 0.90), while maintaining stable convergence and an extremely low risk of overfitting throughout the entire training process. More importantly, the proposed framework can complete end-to-end training and inference on a standard computing environment without GPU acceleration, with a total training time of approximately 16 minutes, substantially enhancing reproducibility, scalability, and deployment potential in resource-limited clinical research settings.

Comparisons with the predefined pure quantum CNN, classical CNN, and ResNet18 baseline models further highlight the distinctive advantages of LQER. Pure quantum models, under the current limited qubit scale, struggle to effectively capture the complex spatial structures present in CAG images and exhibit significantly inferior performance. Purely classical models, although already demonstrating strong discriminative capability, still suffer from inherent limitations in modeling high-order nonlinear feature interactions. In contrast, LQER fully leverages the stability and powerful representational capacity of classical deep networks, while achieving targeted functional enhancement through a small-scale quantum module, forming an efficient collaborative paradigm of ``classical backbone with quantum refinement.'' These empirical findings indicate that, in the current Noisy Intermediate-Scale Quantum (NISQ) era, quantum computing is better positioned as a feature enhancement component for mainstream classical models rather than as a complete replacement. This provides solid theoretical and experimental evidence supporting the application of hybrid quantum--classical learning frameworks in medical image analysis.

From a clinical perspective, LQER achieves highly balanced performance on an independent test set, with an accuracy of 0.950, an AUC of 0.987, and both sensitivity and specificity reaching 0.950, while keeping false positive and false negative rates within 5\%. These results indicate that the model can effectively reduce the risk of missed diagnoses while avoiding excessive false positives, thereby minimizing unnecessary invasive examinations or interventions. The ability to perform prediction based on a single CAG frame confers substantial potential clinical value: the model may serve as a real-time intraoperative decision-support tool to provide objective quantitative references for intermediate stenotic lesions; it may also function as a preliminary screening module prior to multimodal assessment, optimizing diagnostic workflows. Particularly in primary healthcare institutions or resource-constrained environments, LQER can deliver stable and reproducible intelligent assistance, with the potential to significantly improve objectivity and consistency in coronary artery disease diagnosis and treatment decision-making.

Despite the promising results achieved in this study, several limitations and future directions should be acknowledged. First, all experiments were conducted using classical quantum simulators, and validation on real quantum hardware has not yet been performed. Future work should integrate NISQ devices and noise models to investigate performance under more realistic quantum computing conditions. Second, this study focuses on image-level and patch-level binary classification tasks and has not yet been extended to clinically more valuable tasks such as lesion severity grading or continuous risk prediction. Third, the quantum circuit architecture employed remains relatively basic. Future research may explore adaptive quantum circuit depths, variational quantum feature maps, and hybrid enhancement strategies combining quantum modules with attention mechanisms or graph neural networks, to further exploit the potential of quantum computing in high-order feature modeling for medical imaging. Continued investigation along these directions will facilitate the clinical translation and broader application of quantum-enhanced learning frameworks in cardiovascular imaging and beyond.

\section{Conclusion}

This study presents a lightweight quantum-enhanced ResNet (LQER) for coronary angiography image classification, establishing a practical, reproducible, and low-computational-cost hybrid quantum--classical feature enhancement learning framework. Built upon the mature ResNet18 backbone for classical feature extraction, the proposed model introduces a small-scale parameterized quantum circuit into the high-level semantic feature space to selectively enhance critical features, and achieves stable training through end-to-end joint optimization.

Extensive experimental results demonstrate that LQER significantly outperforms both purely classical and purely quantum baseline models across multiple metrics, including accuracy, AUC, F1-score, sensitivity, and specificity, while achieving stable clinical-grade performance in standard computing environments. These findings not only validate the potential value of quantum feature enhancement in medical image analysis, but also provide a feasible technical pathway for translating quantum machine learning from conceptual demonstrations toward practical medical applications.

Overall, this work establishes both methodological and empirical foundations for applying hybrid quantum--classical models to intelligent analysis of coronary angiography, and offers important insights for incorporating quantum computing into more complex medical imaging tasks in future research.

\bibliographystyle{unsrt}
\bibliography{references}

\end{document}